\documentclass{article}
\usepackage[utf8]{inputenc}
\usepackage{a4wide}
\usepackage{xcolor}
\usepackage{amsmath}
\usepackage{lscape}
\usepackage[font=small,labelfont=bf]{caption}
\usepackage{graphicx}
\usepackage{subcaption}
\usepackage{comment}
\usepackage{tabularx}
\usepackage{multirow}
\usepackage{algorithmic}
\usepackage[ruled,vlined]{algorithm2e}
\usepackage{amsmath,amssymb,amsfonts,amsthm} 
{

\theoremstyle{remark}

}
\usepackage[utf8]{inputenc}
\usepackage{amssymb}
\usepackage{multicol}
\usepackage{geometry}
\usepackage{hyperref}
\usepackage{authblk}
\usepackage{ulem}
\usepackage{moreverb}
\usepackage{amsmath,bm}
\usepackage{graphicx}
\usepackage{siunitx}
\usepackage{subcaption}

\renewenvironment{abstract}
  {\small\noindent\textbf{Abstract.}\normalsize\ignorespaces}
  {\par\noindent\ignorespacesafterend}

\makeatletter
\renewcommand{\maketitle}{%
    \noindent\fcolorbox{red}{white}{
        \parbox{\textwidth}{%
            \color{red}Accepted for \textit{The 9th International Symposium on Swarm Behavior and Bio-Inspired Robotics 2025}. 
        }%
    }
    \vspace{1em} 
    \begin{center}%
        {\LARGE \@title \par}
        \vspace{0.5em}
        {\large \@author \par}
        \vspace{0.5em}
        {\large \@date}
    \end{center}%
    \vspace{1em} 
}
\makeatother

\begin{document}

\title{Position-Based Flocking for Robust Alignment}

\author{Hossein B. Jond}
\affil{Department of Cybernetics, Czech Technical University in Prague, Prague, Czechia}

\date{}

\maketitle

\begin{abstract}
This paper presents a position-based flocking model for interacting agents, balancing cohesion-separation and alignment to achieve stable collective motion. The model modifies a position-velocity-based approach by approximating velocity differences using initial and current positions, introducing a threshold weight to ensure sustained alignment. Simulations with 50 agents in 2D demonstrate that the position-based model produces stronger alignment and more rigid and compact formations compared to the position-velocity-based model. The alignment metric and separation distances highlight the efficacy of the proposed model in achieving robust flocking behavior. The model’s use of positions ensures robust alignment, with applications in robotics and collective dynamics.
 
\textbf{Keywords:} alignment, cohesion, flocking, separation. 
\end{abstract}

\section{Introduction}
\label{sec:introduction}

Flocking behavior, observed in natural systems such as bird flocks, fish schools, and insect swarms, represents a fascinating example of emergent collective dynamics \cite{Reynolds1987}. These systems exhibit coordinated motion without centralized control, driven by local interactions among individuals. Understanding and replicating such behaviors in artificial systems have significant implications for robotics, autonomous vehicle coordination, and swarm intelligence \cite{Virágh2014}.

Early models of flocking, such as Reynolds' Boids \cite{Reynolds1987}, introduced three core rules: cohesion (staying close to neighbors), separation (avoiding collisions), and alignment (matching velocities). These principles have inspired numerous mathematical and computational models to capture the dynamics of collective motion \cite{Vicsek1995}, \cite{Cucker2007}, \cite{OlfatiSaber2006}. The Vicsek model, for instance, focuses on alignment through local velocity averaging, demonstrating phase transitions from disordered to ordered motion. However, such models often rely on velocity information, which is difficult to measure directly in robotic systems or requires error-prone estimation due to sensor noise \cite{Schilling2021}, \cite{Wang2024}. Thus, flocking based solely on position data is both practical and significant for robust coordination.

Position-based flocking models have emerged to address these limitations, using relative positions to infer interaction rules, thus reducing dependence on velocity data \cite{Zhan2013}. These models are particularly relevant for systems where position sensing is more feasible, such as in GPS-equipped drones or sensor networks \cite{Virágh2014}. Position-based approaches often incorporate potential functions to balance attraction and repulsion, ensuring stable formations. These approaches are preferred in robotic multi-agent systems (MAS) for their simplicity, robustness, and efficiency in both perception-based and communication-based systems. They leverage direct position measurements to minimize computational and communication demands while enhancing scalability and fault tolerance.

Alignment remains a critical component of flocking, as it drives velocity consensus among agents \cite{Cucker2007}. Position-based alignment rules, which approximate velocity differences using sampled position data, enhance robustness in noisy environments and reduce communication bandwidth.  For second-order multi-agent systems, position-based protocols have been shown to achieve consensus by utilizing sampled or relative position data, even in the absence of direct velocity measurements or communication \cite{Mei2013}, \cite{Yu2011}. Time-varying topologies further complicate consensus, yet position-based rules can ensure coordination under strongly connected networks \cite{Gao2009}.

This paper builds on prior work, proposing a position-based flocking model that modifies the position-velocity-based framework of \cite{jond2025}. Like its position-velocity-based counterpart \cite{jond2025}, our model introduces a minimal, distributed, and scalable position-based approach for robotic MAS. By approximating velocity differences using initial and current positions and introducing a threshold weight, the model ensures robust alignment and compact formations. The following sections detail the model formulation, numerical simulations, and a comparison with position-velocity-based approaches, demonstrating enhanced performance in achieving stable flocking behavior.

\section{Flocking from Position and Velocity States}
Consider \(n \geq 2\) agents in an \(m\)-dimensional space, with positions \(\mathbf{p}_i\) and velocities \(\mathbf{v}_i\). The neighborhood of agent \(i\) is \(\mathcal{N}_i = \{j \neq i \mid \|\mathbf{p}_j - \mathbf{p}_i\| \leq r_i\}\), where \(r_i > 0\) is the interaction radius. The flocking model from \cite{jond2025} is
\begin{align}\label{eq:velocity-model}
    \frac{d\mathbf{p}_i}{dt} &= \mathbf{v}_i, \notag\\
    \frac{d\mathbf{v}_i}{dt} &= \sum_{j \in \mathcal{N}_i} \psi(\|\mathbf{p}_j - \mathbf{p}_i\|)(\mathbf{p}_j - \mathbf{p}_i) + \sum_{j \in \mathcal{N}_i} \phi (\mathbf{v}_j - \mathbf{v}_i),
\end{align}
where \(\psi(\|\mathbf{p}_j - \mathbf{p}_i\|) = 1 - \frac{|\mathcal{N}_i|}{\|\mathbf{p}_j - \mathbf{p}_i\|}\), \(\phi = \frac{1}{|\mathcal{N}_i|}\), are the interaction weights and \(|\mathcal{N}_i|\) is the neighborhood size. Velocity is constrained by
\[
\mathbf{v}_i = v_i^{\max} \tanh\left(\frac{\|\mathbf{v}_i\|}{v_i^{\max}}\right) \frac{\mathbf{v}_i}{\|\mathbf{v}_i\|},
\]
with maximum speed \(v_i^{\max}\) and saturation \(s_i = \frac{v_i^{\max}}{t_i^{v_{\max}}}\) for the velocity rate, with \(t_i^{v_{\max}}\) the time to reach \( v_i^{\max} \).

The model balances cohesion-separation and alignment. The cohesion-separation (aggregation) potential combines short-range repulsion and long-range attraction, setting the equilibrium spacing at \(|\mathcal{N}_i|\). Smaller \(|\mathcal{N}_i|\) yields denser formations, larger \(|\mathcal{N}_i|\) looser ones, ensuring collision avoidance. The alignment term drives velocity consensus.

\section{Position-Based Flocking}\label{sec:position-based-flocking}

At \(t \to \infty\), \(\sum_{j \in \mathcal{N}_i} (\mathbf{v}_j - \mathbf{v}_i) \to 0\), implying parallel velocities and stable relative positions, as \(\frac{d}{dt} \sum_{j \in \mathcal{N}_i} (\mathbf{p}_j - \mathbf{p}_i) \to 0\). Using the derivative definition, the alignment term is
\[
\sum_{j \in \mathcal{N}_i} \phi (\mathbf{v}_j - \mathbf{v}_i) = \frac{d}{dt} \sum_{j \in \mathcal{N}_i} \phi(\mathbf{p}_j - \mathbf{p}_i).
\]
For a position-based rule, approximate average velocity over \([0, t]\)
\[
\mathbf{v}_j - \mathbf{v}_i \approx \frac{(\mathbf{p}_j - \mathbf{p}_i) - (\mathbf{p}_j(0) - \mathbf{p}_i(0))}{t}.
\]
Thus, the alignment term becomes
\[
\sum_{j \in \mathcal{N}_i} \phi (\mathbf{v}_j - \mathbf{v}_i) \approx  \sum_{j \in \mathcal{N}_i} \frac{\phi}{t}\left[ (\mathbf{p}_j - \mathbf{p}_i) - (\mathbf{p}_j(0) - \mathbf{p}_i(0)) \right].
\]
The weight \(\frac{\phi}{t}\) drives alignment effectively at small \(t\), but diminishes as \(t \to \infty\). To ensure sustained alignment, we introduce a threshold weight \( \hat{\phi}_{\text{min}} = k  |\mathcal{N}_i| \), with \( k > 0 \) (e.g., \( k = 0.1 \) for \( n = 50 \), \( r_i = 7.5\, \text{m} \)). The modified position-based alignment rule is
\begin{equation*}
      \sum_{j \in \mathcal{N}_i} \max\big( \frac{|\mathcal{N}_i|}{t},  \hat{\phi}_{\text{min}}\big)\left[ (\mathbf{p}_j - \mathbf{p}_i) - (\mathbf{p}_j(0) - \mathbf{p}_i(0)) \right],
\end{equation*}

This rule scales the influence of neighbors based on local density, promoting faster convergence for $t < \frac{1}{k}$ while preserving alignment for $t \geq \frac{1}{k}$ using the alignment strength at $t = \frac{1}{k}$. It adjusts agent $i$’s velocity to minimize changes in relative positions from $t = 0$, thereby promoting velocity alignment over time using only the initial and current positions.

The resulting position-based flocking model is
\begin{align}\label{eq:position-model}
    \frac{d\mathbf{p}_i}{dt}&=\mathbf{v}_i, \notag\\
    \frac{d\mathbf{v}_i}{dt}&=\sum_{j\in\mathcal{N}_i}\hat{\psi}(\lVert \mathbf{p}_j-\mathbf{p}_i\rVert)(\mathbf{p}_j-\mathbf{p}_i)\notag\\&+\sum_{j\in\mathcal{N}_i}\hat{\phi}(\mathbf{p}_j(0)-\mathbf{p}_i(0)),
\end{align}
with weights \(\hat{\psi}(\|\mathbf{p}_j - \mathbf{p}_i\|) = \psi(\|\mathbf{p}_j - \mathbf{p}_i\|) - \hat{\phi}\) and \(\hat{\phi} = -\max\big( \frac{|\mathcal{N}_i|}{t}, \hat{\phi}_{\text{min}}\big)\). The influence of initial positions diminishes when \( t \geq \frac{1}{k} \), stabilizing at \( \hat{\phi} = \hat{\phi}_{\text{min}} \).

For robotic MAS without direct communication, the position-based model \eqref{eq:position-model} offers simpler sensing (e.g., LIDAR), robustness to noise, and lower processing demands compared to the position-velocity-based model \eqref{eq:velocity-model}. With communication, it reduces bandwidth, simplifies data processing, and enhances fault tolerance by using last-known positions. Intrinsically, the communication cost is reduced by half as a result of the reliance on position-only measurements.

\section{Numerical Simulations} \label{sec:numerical-simulations}

To evaluate  the position-based flocking models \eqref{eq:position-model}, we simulate \(n = 50\) agents in a 2D space (\(m = 2\)) over \(t \in [0, 100]\, \text{s}\). Initial positions \(\mathbf{p}_i(0)\) are randomly distributed in a \(25 \times 25\, \text{m}\) square, with initial velocities \(\mathbf{v}_i(0)\) randomly oriented, satisfying \(\|\mathbf{v}_i(0)\| \leq1\, \text{m/s}\). The interaction radius is \(r_i = 7.5\, \text{m}\), \( v_i^{\max} = 5\, \text{m/s}\), and \(t_i^{v_{\max}} = 1\, \text{s}\), so \(s_i = 5\; \mathrm{m/s^2}\). 

We measure directional alignment using the metric
\[
\gamma = \frac{1}{n}\sum_{i=1}^n \frac{1}{|\mathcal{N}_i|}\sum_{j \in \mathcal{N}_i}\frac{\mathbf{v}_i^\top \mathbf{v}_j}{\|\mathbf{v}_i\|\|\mathbf{v}_j\|},
\]
where the inner term computes the average cosine similarity between agent \( i \)’s velocity and its neighbors’ velocities, and the outer term averages across all agents. The metric \( \gamma \in [-1, 1] \), with \( \gamma \approx 1 \) indicating near-parallel velocities (strong alignment) and values near or below 0 indicating misalignment. By normalizing direction, this metric isolates directional consensus from speed differences.

\begin{figure}
    \centering
    \begin{subfigure}{0.2\textwidth}
        \centering
        \includegraphics[width=\linewidth]{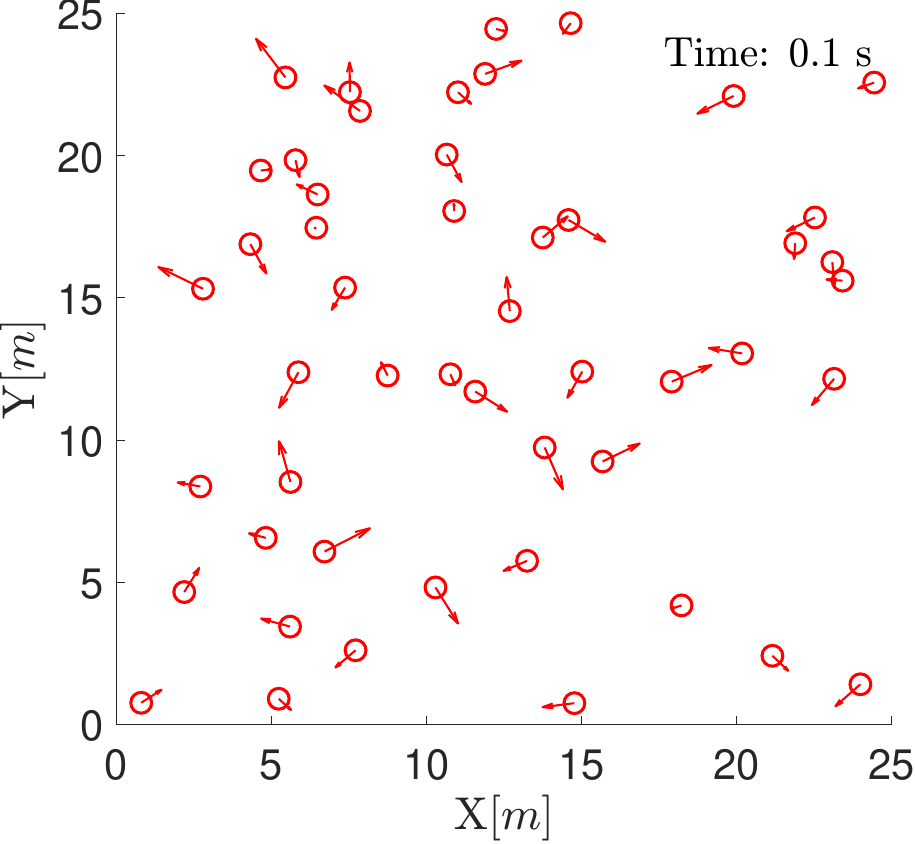}
        \caption{}
    \end{subfigure}
        \begin{subfigure}{0.2\textwidth}
        \centering
        \includegraphics[width=\linewidth]{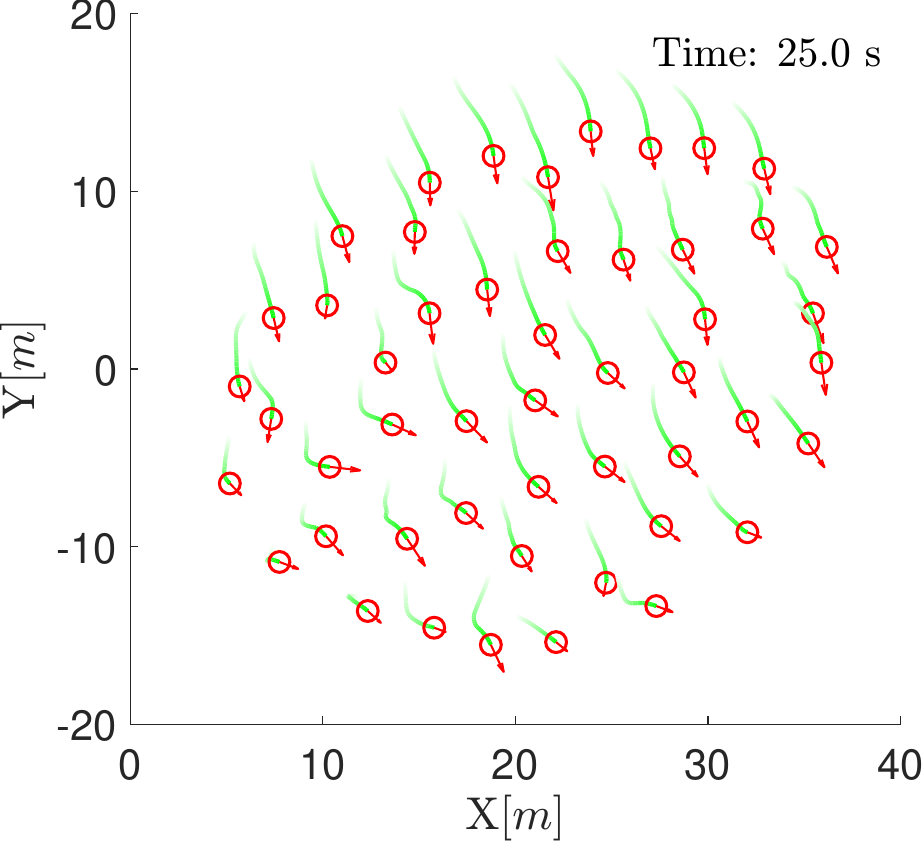}
        \caption{}
    \end{subfigure}
    \begin{subfigure}{0.2\textwidth}
        \centering
        \includegraphics[width=\linewidth]{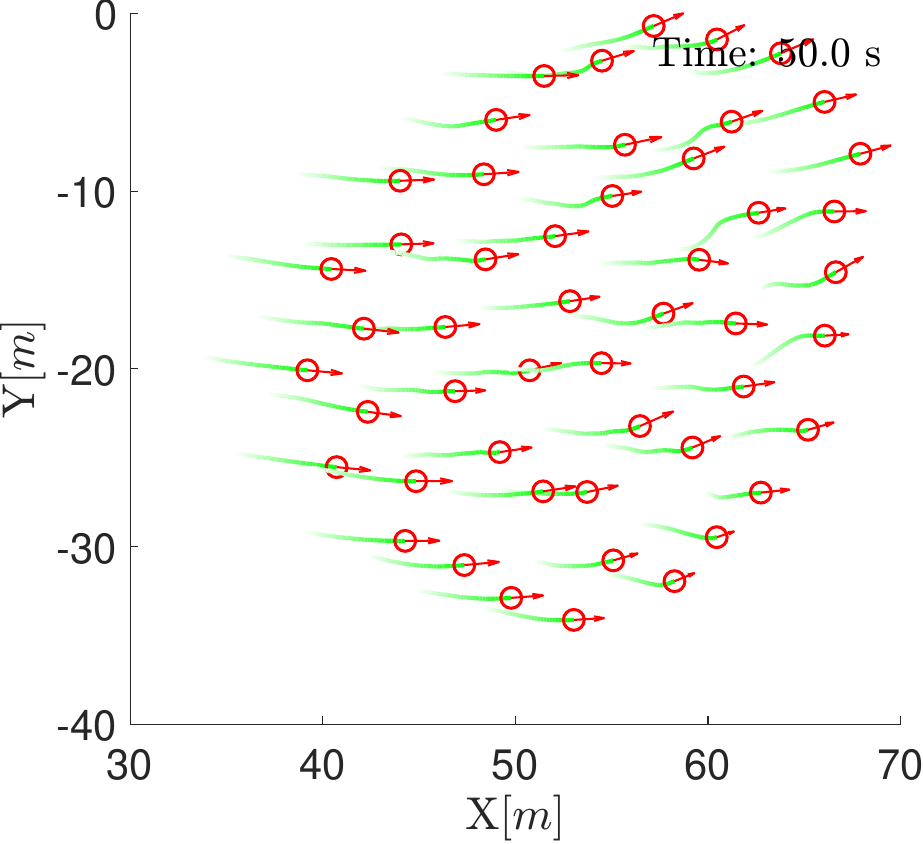}
        \caption{}
    \end{subfigure}
    \begin{subfigure}{0.2\textwidth}
        \centering
        \includegraphics[width=\linewidth]{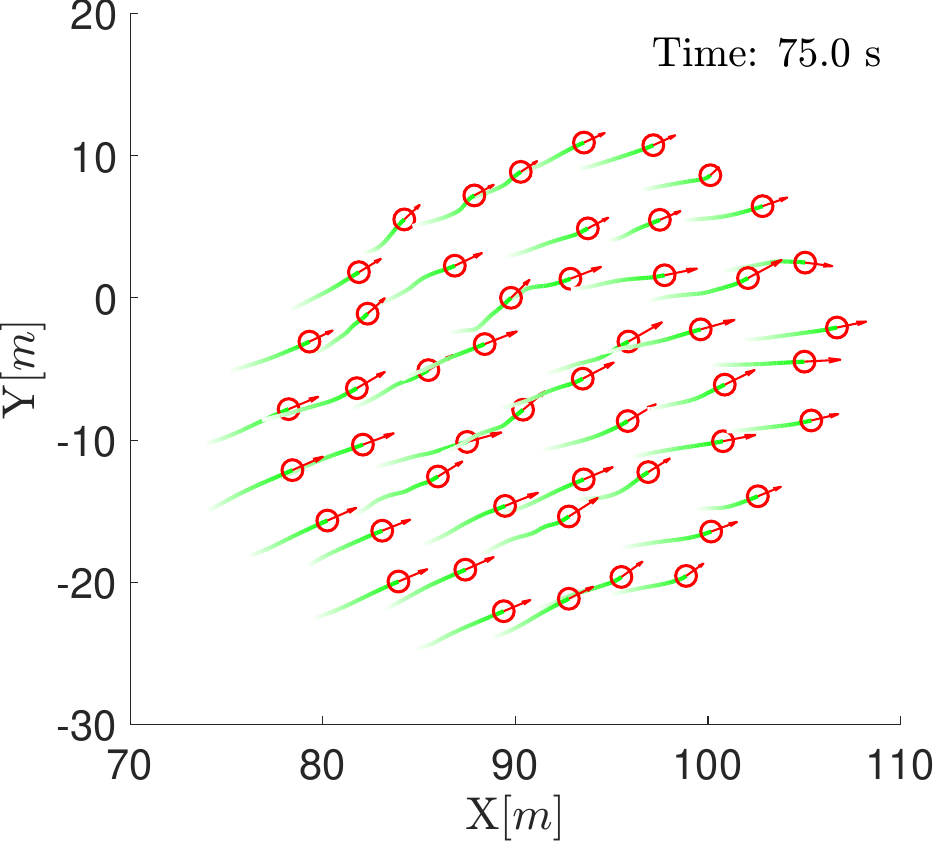}
        \caption{}
    \end{subfigure}
    \begin{subfigure}{0.2\textwidth}
        \centering
        \includegraphics[width=\linewidth]{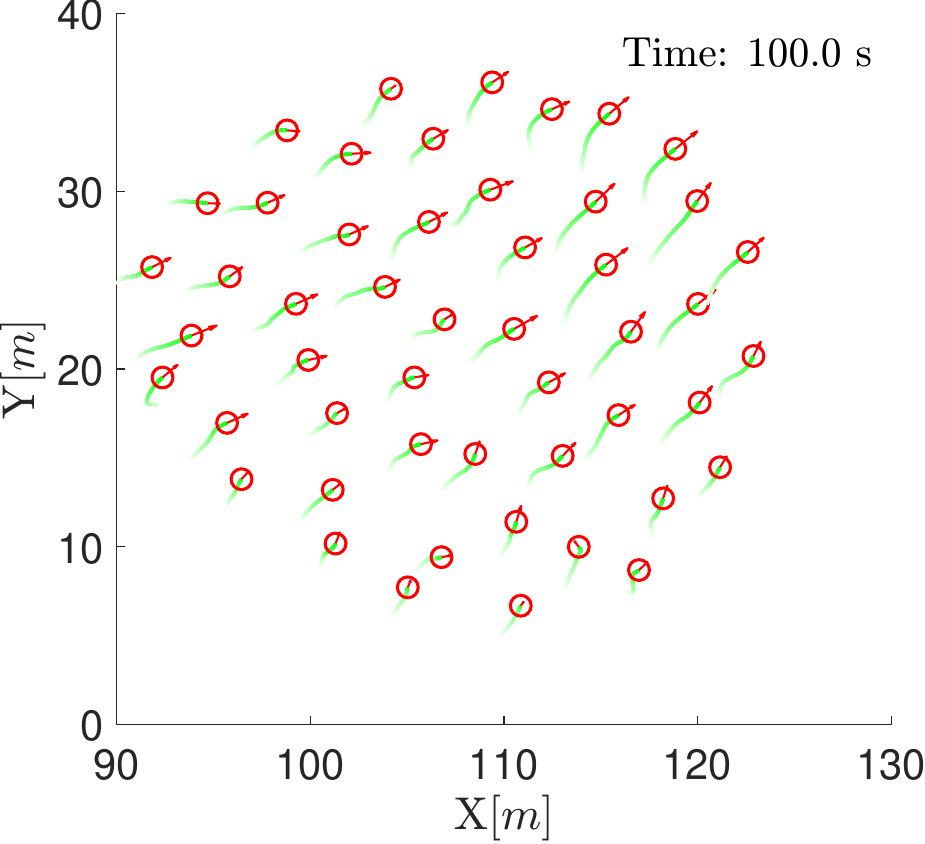}
        \caption{}
    \end{subfigure} 
    \caption{Snapshots of flocking behavior under the position-velocity-based model \eqref{eq:velocity-model} at \(t=0, 25, 50, 75, 100 \, \text{s}\). From (a) random initial conditions, (b) shows partial alignment, (c) and (d) approach full flocking, and (e) shows partial alignment in the final configuration. The red arrow represents the velocity vector of the agent, with its length denoting the agent’s speed, and the green trajectory indicates the agent’s path over the last 3 s. }
    \label{fig:flock-vel}
\end{figure}

\begin{figure}
    \centering
    \begin{subfigure}{0.22\textwidth}
        \centering
        \includegraphics[width=\linewidth]{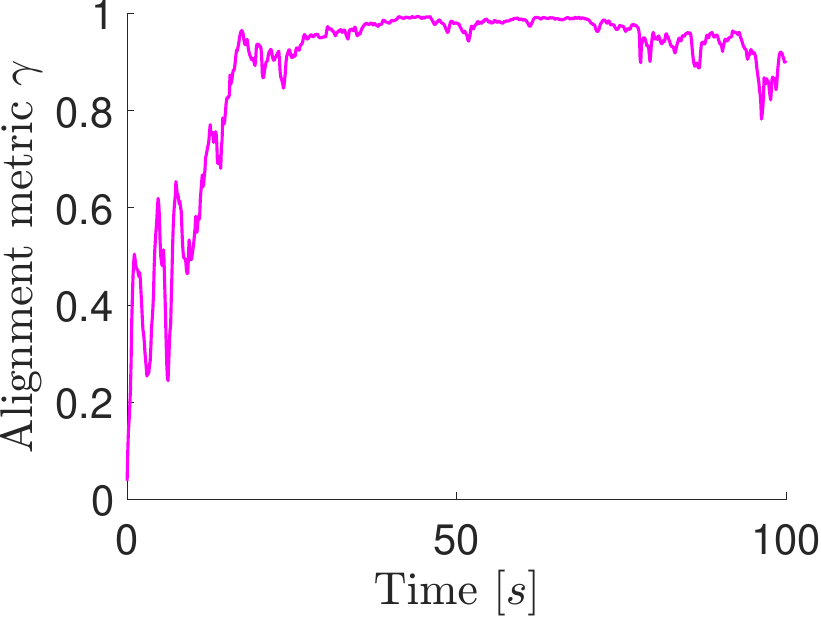}
        \caption{}
    \end{subfigure}
        \begin{subfigure}{0.22\textwidth}
        \centering
        \includegraphics[width=\linewidth]{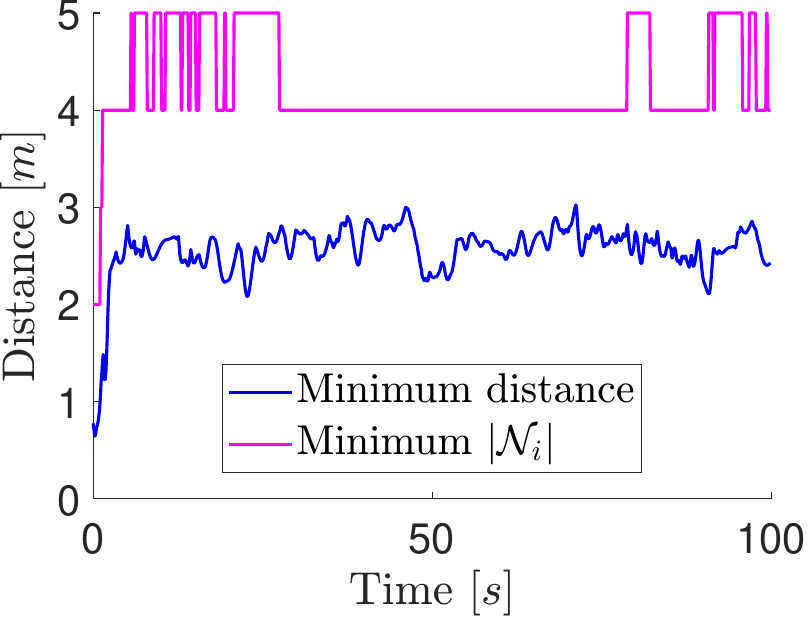}
        \caption{}
    \end{subfigure}
    \caption{Time histories for the position-velocity-based model \eqref{eq:velocity-model}: (a) alignment metric \( \gamma \), (b) minimum separation distance and neighborhood size  \(|\mathcal{N}_i|\).}
    \label{fig:flock-vel-hist}
\end{figure}

Fig. \ref{fig:flock-vel} shows snapshots of flocking behavior under the position-velocity-based model \eqref{eq:velocity-model} at \( t = 0, 25, 50, 75, 100 \, \text{s} \). From (a) random initial conditions, (b) shows partial alignment, (c) and (d) approach full flocking, and (e) indicates a final configuration with partial alignment.  Fig. \ref{fig:flock-vel-hist} tracks (a) the alignment metric $\gamma$ and (b) the minimum separation distance and neighborhood size $|\mathcal{N}_i|$ over time. Peak alignment occurs between \( t = 20 \, \text{s} \) and \( t = 80 \, \text{s} \), with a decline after \( t = 80 \, \text{s} \), consistent with the partially aligned final state.

\begin{figure}
    \centering
    \begin{subfigure}{0.2\textwidth}
        \centering
        \includegraphics[width=\linewidth]{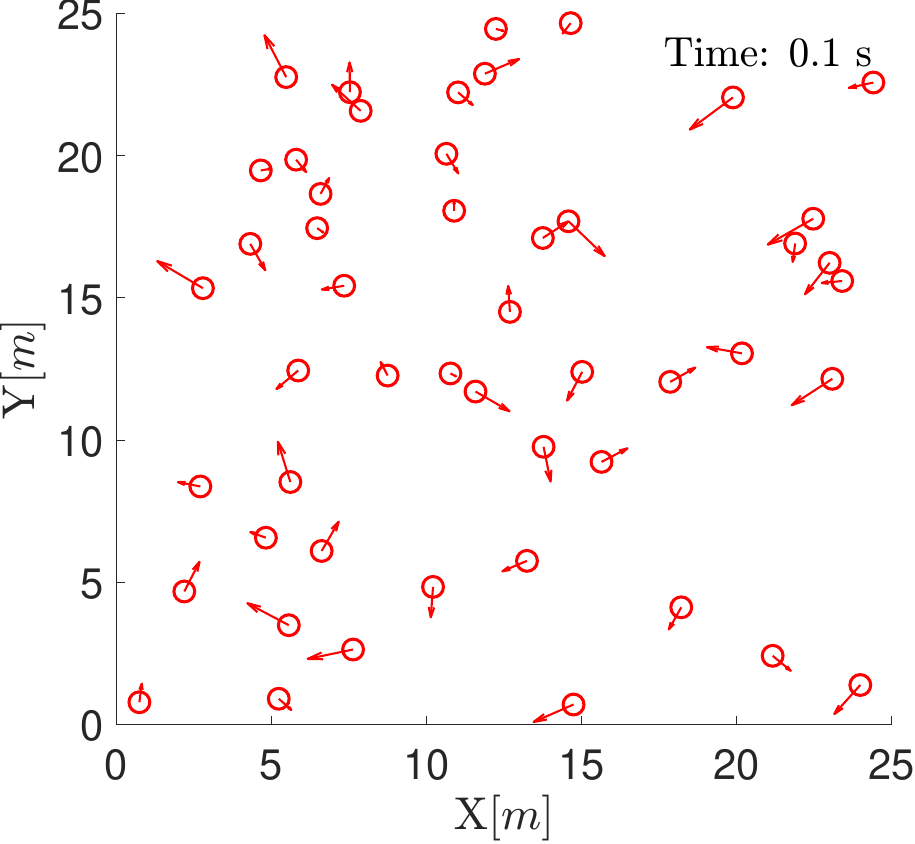}
        \caption{}
    \end{subfigure}
        \begin{subfigure}{0.2\textwidth}
        \centering
        \includegraphics[width=\linewidth]{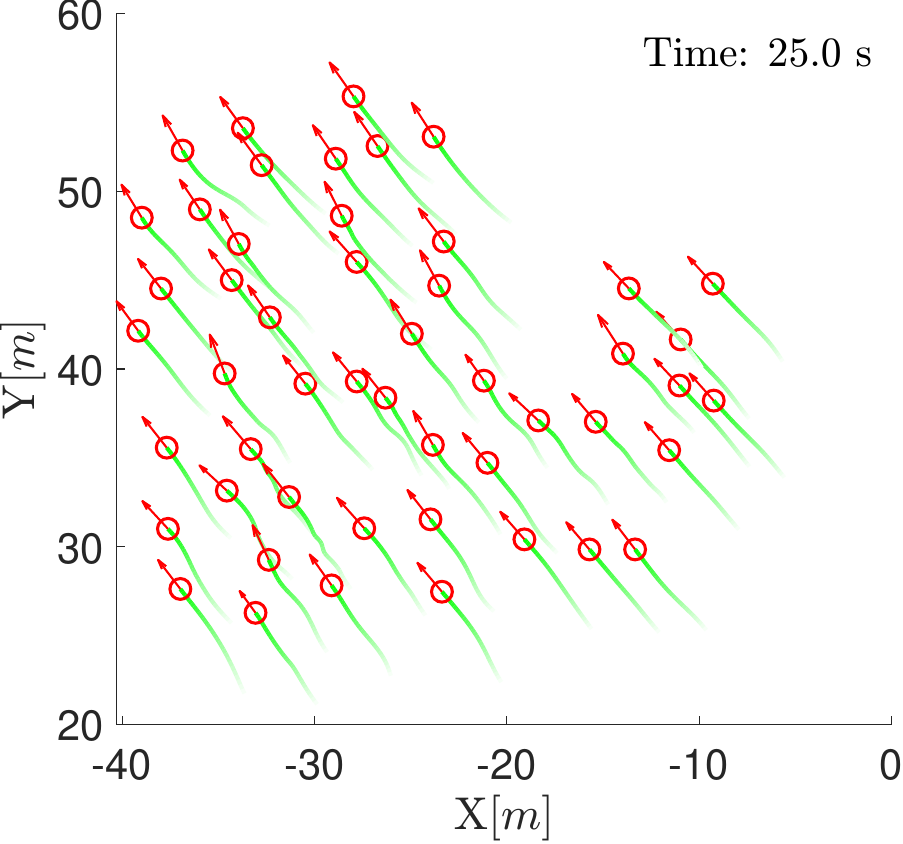}
        \caption{}
    \end{subfigure}
    \begin{subfigure}{0.2\textwidth}
        \centering
        \includegraphics[width=\linewidth]{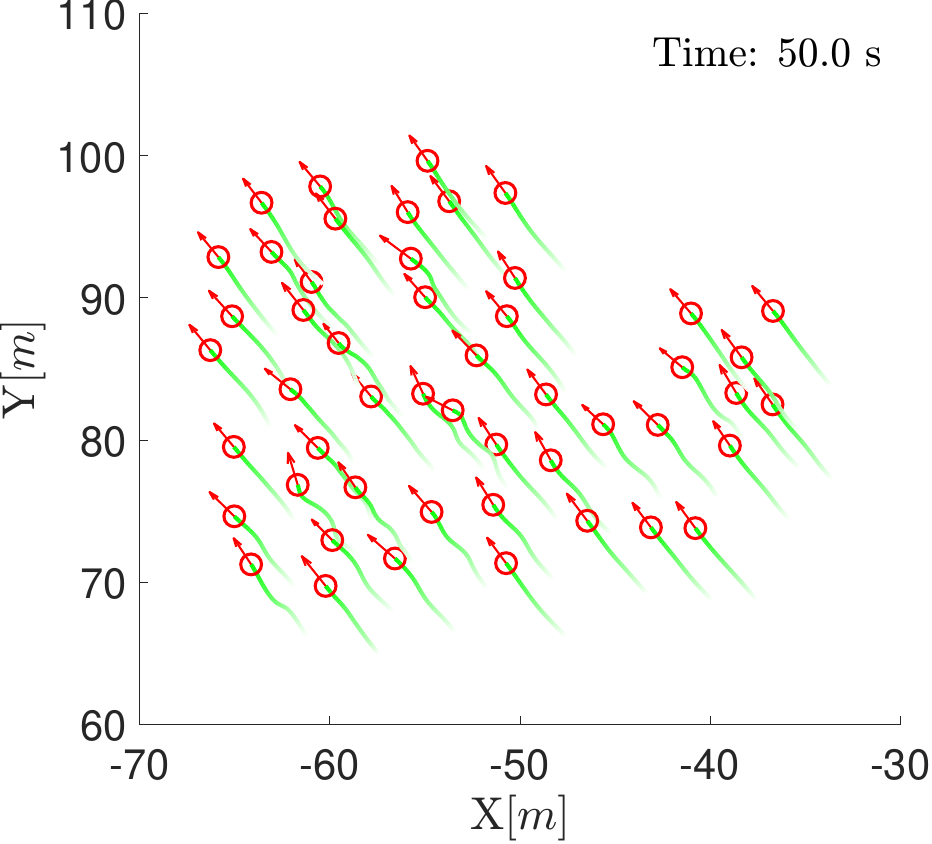}
        \caption{}
    \end{subfigure}
    \begin{subfigure}{0.2\textwidth}
        \centering
        \includegraphics[width=\linewidth]{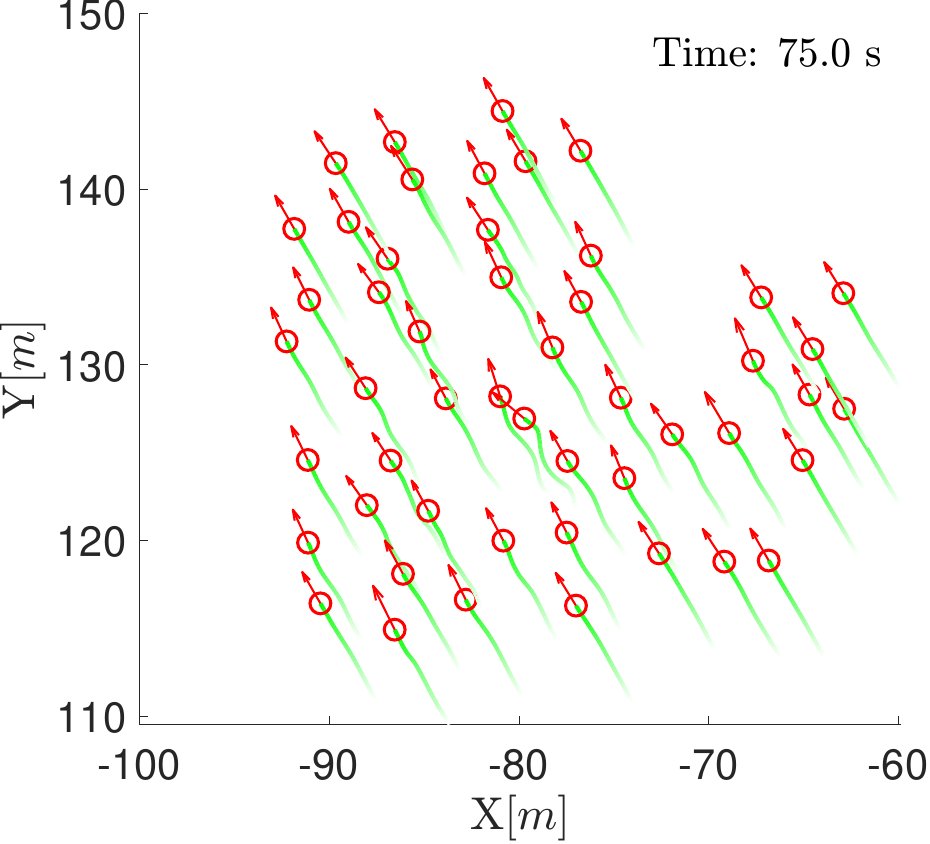}
        \caption{}
    \end{subfigure}
    \begin{subfigure}{0.2\textwidth}
        \centering
        \includegraphics[width=\linewidth]{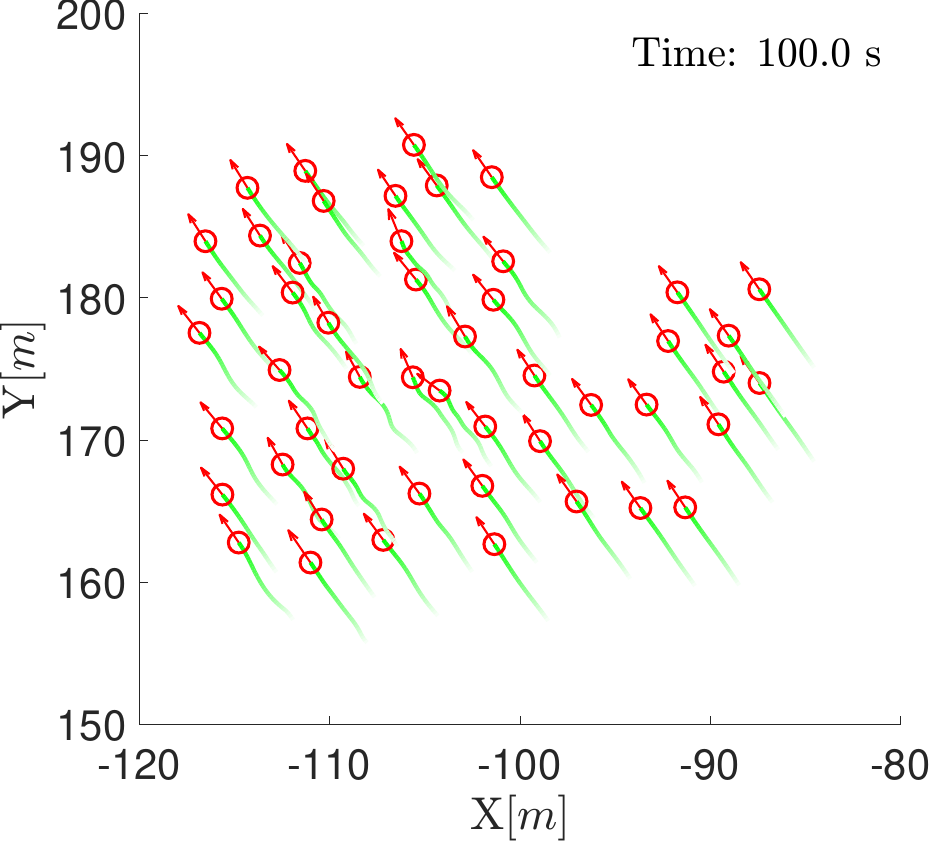}
        \caption{}
    \end{subfigure} 
 \caption{Snapshots of flocking behavior under the position-based model \eqref{eq:position-model} at \(t=0, 25, 50, 75, 100 \, \text{s}\). From (a) random initial conditions, (b) to (e) show strong alignment and a rigid formation.}
    \label{fig:position-flocking}
\end{figure}

\begin{figure}
    \centering
    \begin{subfigure}{0.22\textwidth}
        \centering
        \includegraphics[width=\linewidth]{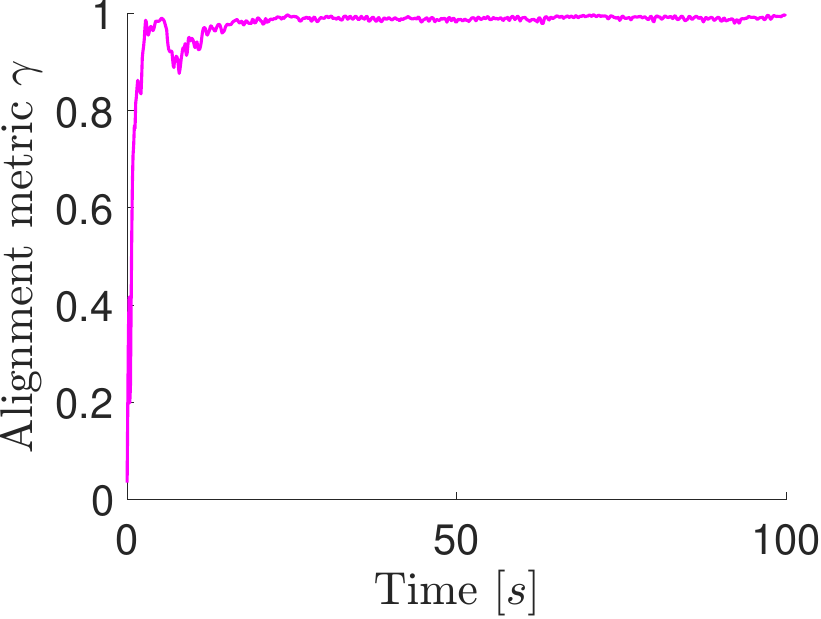}
        \caption{}
    \end{subfigure}
        \begin{subfigure}{0.22\textwidth}
        \centering
        \includegraphics[width=\linewidth]{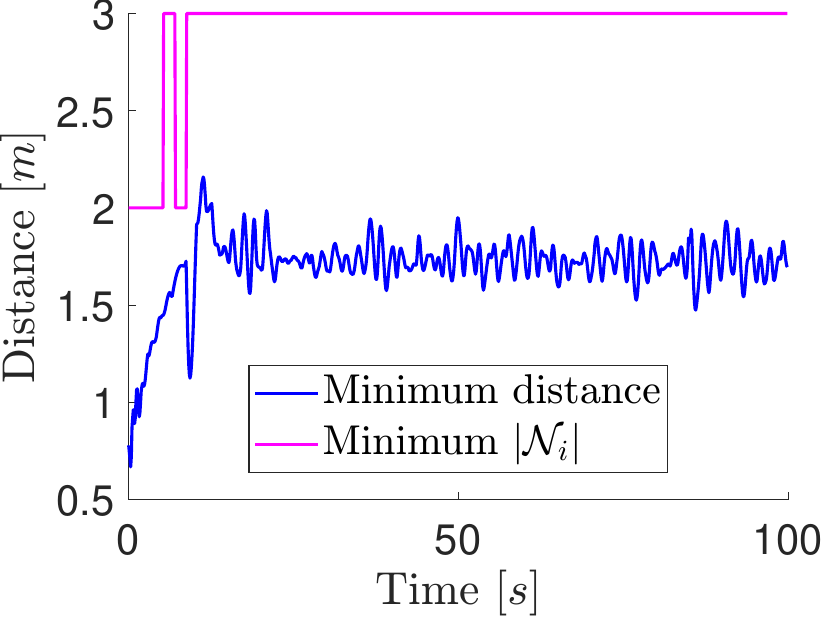}
        \caption{}
    \end{subfigure}
    \caption{Time histories for the position-based model \eqref{eq:position-model}: (a) alignment metric \( \gamma \), (b) minimum separation distance and neighborhood size \( |\mathcal{N}_i| \).}
    \label{fig:position-metrics}
\end{figure}

Fig.~\ref{fig:position-flocking} illustrates flocking under the position-based model \eqref{eq:position-model} at the same time points. From (a) identical initial conditions, snapshots (b) to (e) show strong alignment, forming a rigid flocking configuration. Fig.~\ref{fig:position-metrics} tracks (a) the alignment metric \( \gamma \), showing rapid convergence to strong alignment within 10 s, sustained throughout, and (b) the minimum separation distance (1.5--2 m) and neighborhood size \( |\mathcal{N}_i| \). Compared to the position-velocity-based model (separation 2--3 m), the position-based model yields a more rigid and compact formation.

\begin{figure}
    \centering
    \begin{subfigure}{0.2\textwidth}
        \centering
        \includegraphics[width=\linewidth]{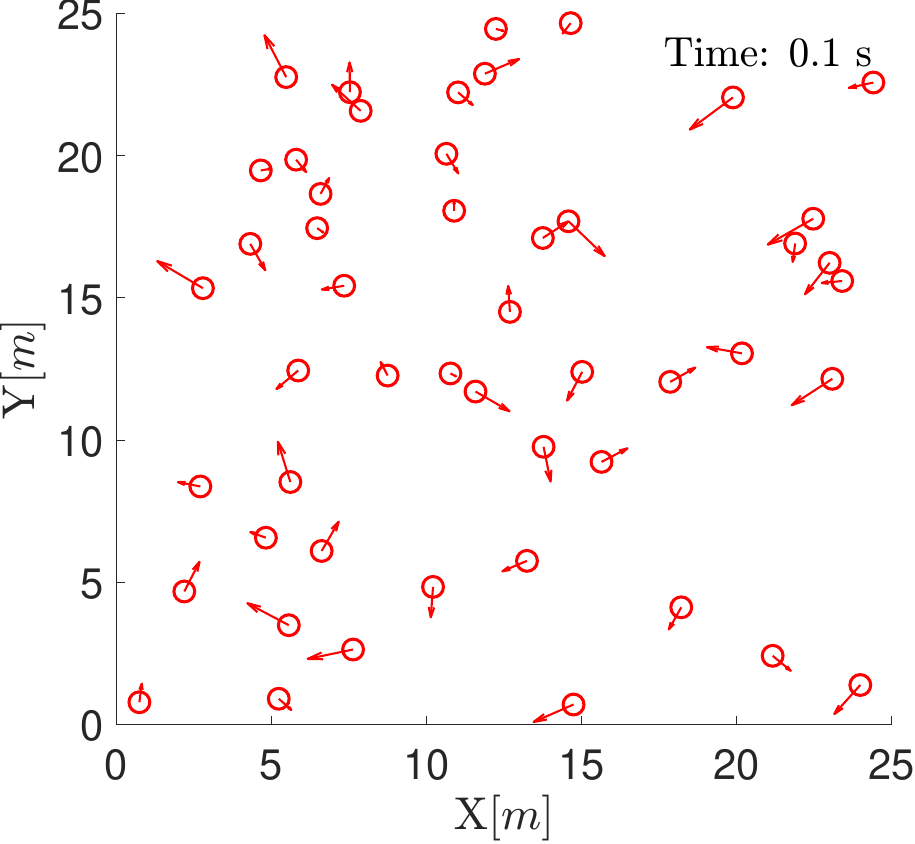}
        \caption{}
    \end{subfigure}
        \begin{subfigure}{0.2\textwidth}
        \centering
        \includegraphics[width=\linewidth]{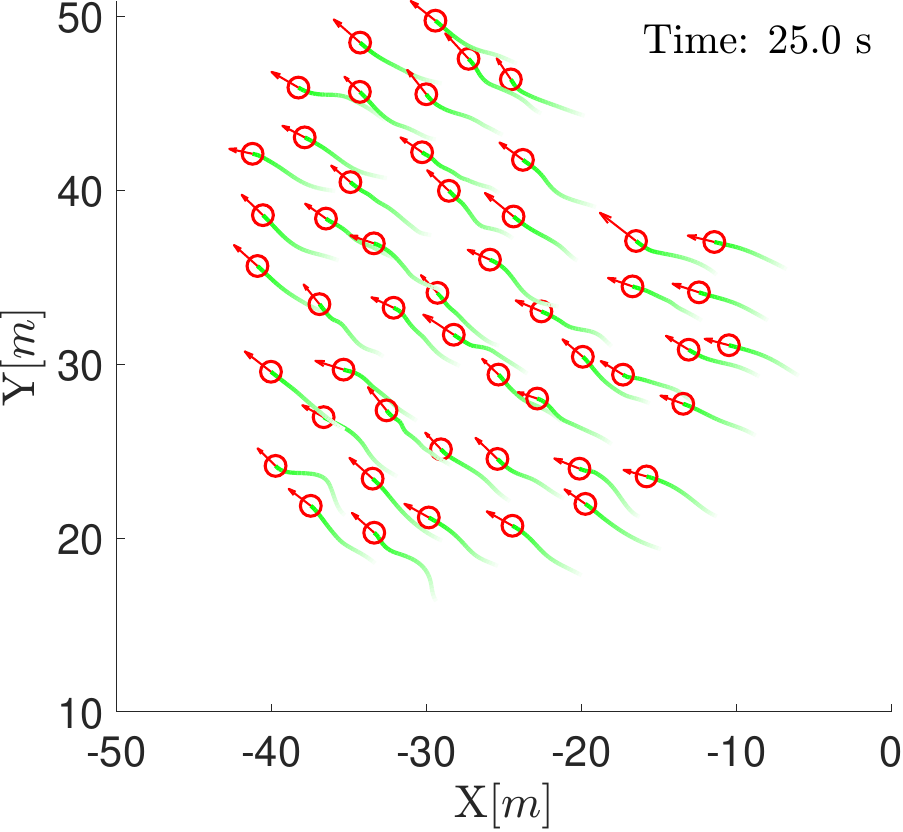}
        \caption{}
    \end{subfigure}
    \begin{subfigure}{0.2\textwidth}
        \centering
        \includegraphics[width=\linewidth]{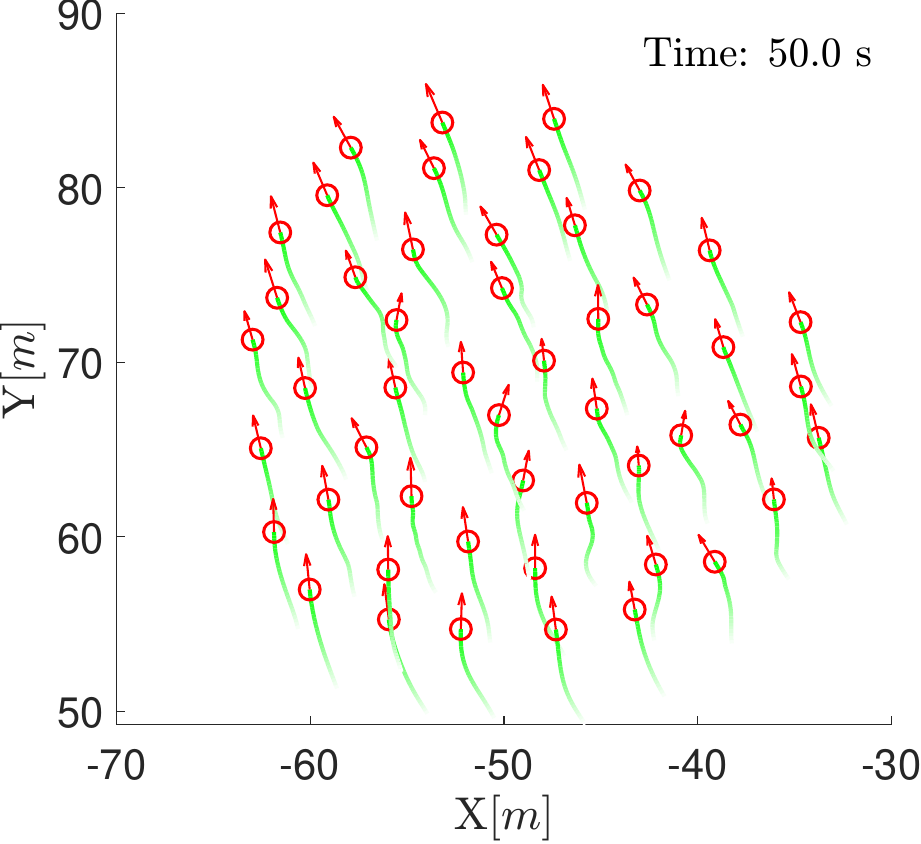}
        \caption{}
    \end{subfigure}
    \begin{subfigure}{0.2\textwidth}
        \centering
        \includegraphics[width=\linewidth]{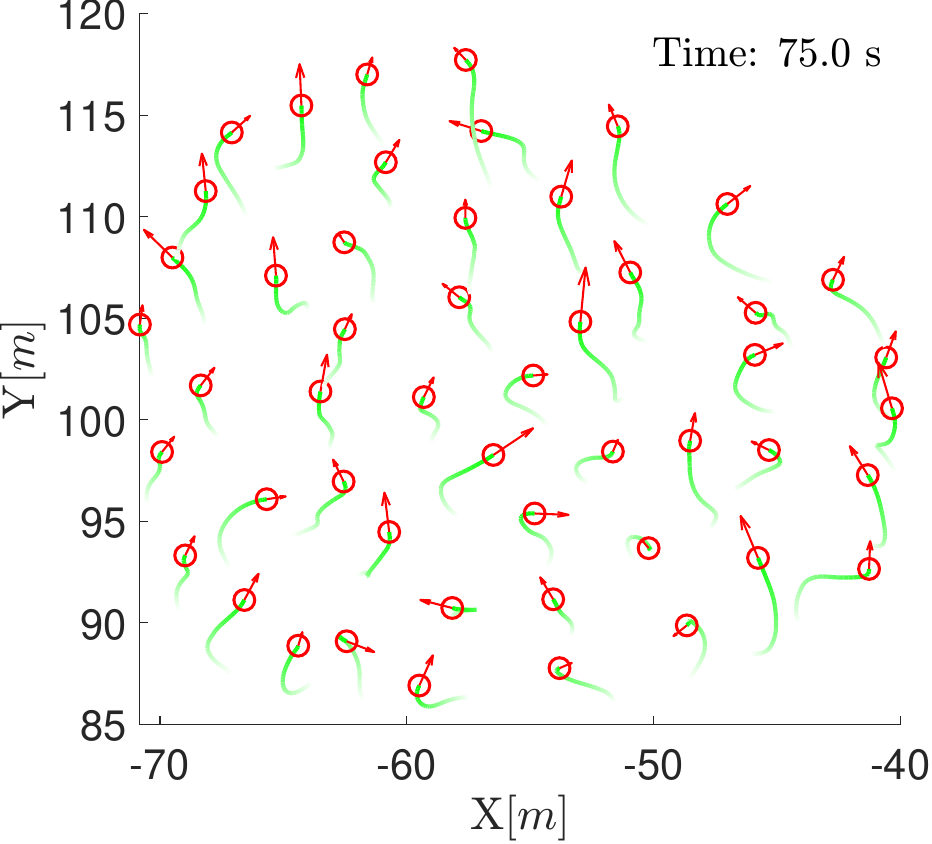}
        \caption{}
    \end{subfigure}
    \begin{subfigure}{0.2\textwidth}
        \centering
        \includegraphics[width=\linewidth]{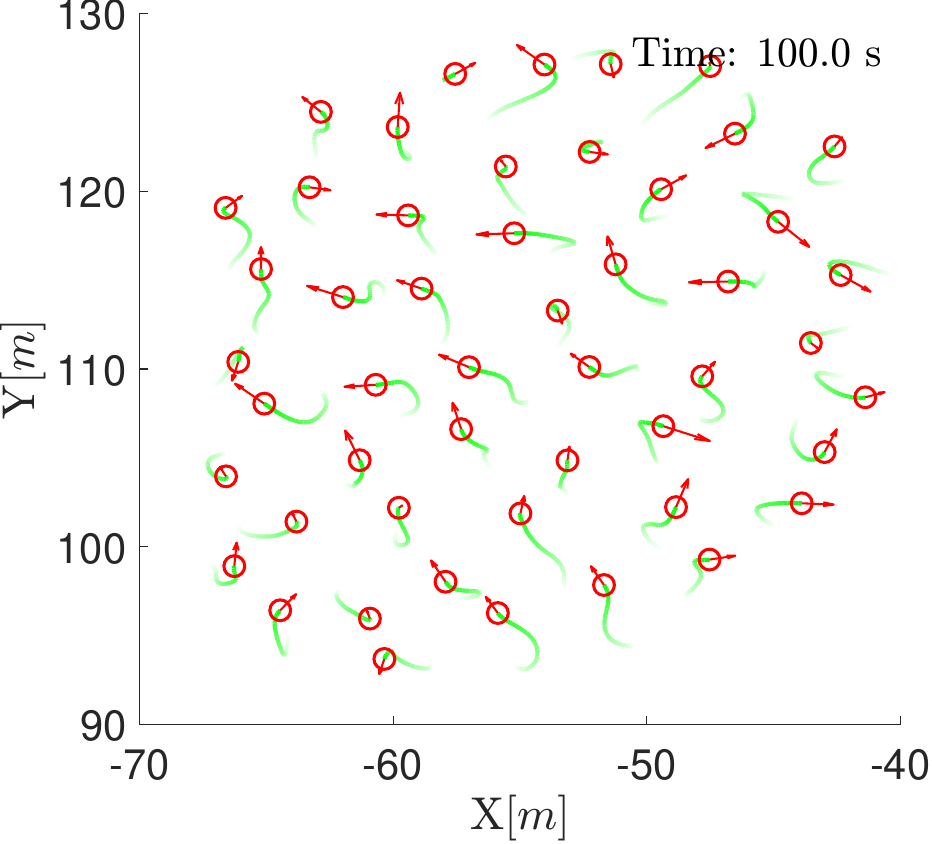}
        \caption{}
    \end{subfigure} 
    \caption{Snapshots of flocking behavior under the position-based model \eqref{eq:position-model} without threshold weight (\( \hat{\phi} = \frac{|\mathcal{N}_i|}{t} \)). From (a) random initial conditions, (b) and (c) show strong alignment, but (d) and (e) indicate loss of alignment.}
    \label{fig:position-flocking-no-threshold}
\end{figure}

\begin{figure}
    \centering
    \begin{subfigure}{0.22\textwidth}
        \centering
        \includegraphics[width=\linewidth]{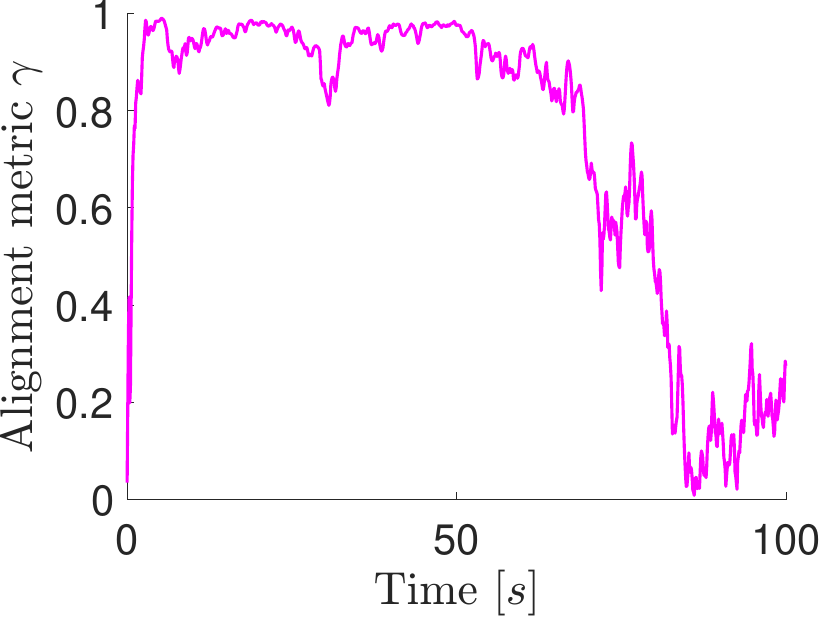}
        \caption{}
    \end{subfigure}
        \begin{subfigure}{0.22\textwidth}
        \centering
        \includegraphics[width=\linewidth]{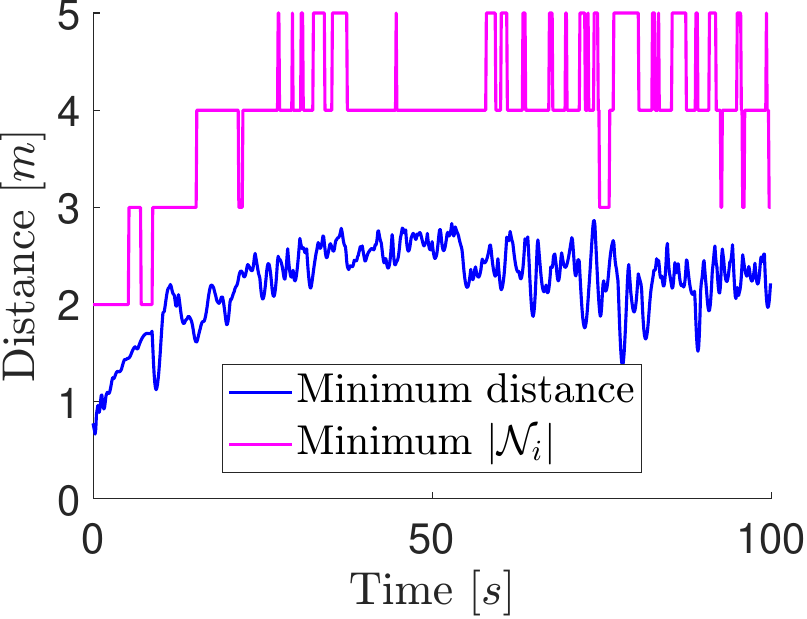}
        \caption{}
    \end{subfigure}
    \caption{Time histories for the position-based model \eqref{eq:position-model} without threshold weight: (a) alignment metric \( \gamma \), (b) minimum separation distance and neighborhood size \( |\mathcal{N}_i| \).}
    \label{fig:position-metrics-no-threshold}
\end{figure}

To demonstrate the importance of the threshold weight \( \hat{\phi}_{\text{min}} \), we simulate the position-based model with \( \hat{\phi} = \frac{|\mathcal{N}_i|}{t} \). Fig.~\ref{fig:position-flocking-no-threshold} shows that, while (b) and (c) exhibit strong alignment, (d) and (e) indicate loss of alignment due to the decaying weight. Fig.~\ref{fig:position-metrics-no-threshold} confirms this, with (a) showing declining \( \gamma \) as \( t \) increases, and (b) tracking separation and \( |\mathcal{N}_i| \). The threshold \( \hat{\phi}_{\text{min}} \) ensures sustained alignment, as seen in Figs.~\ref{fig:position-flocking} and \ref{fig:position-metrics}.

Table~\ref{tab:flocking-comparison} summarizes the characteristics of the position-velocity-based (Pos-Vel) and position-based models.

\begin{table}
\caption{Comparison of flocking models characteristics.}
\label{tab:flocking-comparison}
    \centering
    \begin{tabular}{|l|c|c|}
        \hline
        \textbf{Characteristic} & \textbf{Pos-Vel-Based} & \textbf{Position-Based} \\
        \hline
        Alignment & Strong & Very strong \\
        \hline
        Formation & Flexible & Rigid \\
        \hline
       Separation & Wider & Narrower \\
       \hline
    \end{tabular}
\end{table}

\section{Conclusion}
\label{sec:conclusion}

This study presents a position-based flocking model that approximates velocity differences using initial and current positions with a time- and density-dependent weight. To counter the weakening of alignment from diminishing weights, a threshold weight ensures sustained coordination. Simulations show the model achieves stronger alignment and more compact formations than its position-velocity-based counterpart. By leveraging position data and a threshold weight, it ensures robust velocity consensus and stable spatial arrangements. Alignment metrics and separation distances validate the model’s effectiveness in cohesive flocking. Future work could explore adaptive interaction radii, noise robustness, or dynamic thresholds to improve flexibility and scalability in diverse flocking scenarios.

\section*{Acknowledgments} This work was funded by the Czech Science Foundation (GAČR) under research project no. \(\mathrm{23-07517S}\) and the European Union under the project Robotics and Advanced Industrial Production (reg. no. \(\mathrm{CZ.02.01.01/00/22\_008/0004590}\)).


\end{document}